\documentclass[runningheads]{llncs}

\usepackage{amsmath}
\usepackage{amssymb}
\usepackage{theorem}
\usepackage{graphicx}
\usepackage{hyperref}

\def\PV{\mathcal{P}}
\def\QV{\mathcal{Q}}
\def\PVW{\mathcal{P}_{w}}

\def\pp{\ldotp\ldotp}

\def\cd3#1{\textbf{\textsf{#1}}}
\def\sa#1{\cd3{#1}}

\def\lp{\textit{lp}}

\begin{document}

\title{Abelian Repetitions in Sturmian Words}

\author{Gabriele Fici\inst{1}, Alessio Langiu\inst{2}, Thierry Lecroq\inst{3}, Arnaud Lefebvre\inst{3},  Filippo~Mignosi\inst{4} and \'Elise Prieur-Gaston\inst{3}}

\authorrunning{G. Fici et al.}

\institute{
Dipartimento di Matematica e Informatica, Universit\`a di Palermo, Italy\\ \email{Gabriele.Fici@unipa.it}  \and  
Department of Informatics, King's College London, London, UK \\
\email{Alessio.Langiu@kcl.ac.uk} \and
Normandie Universit\'e, LITIS EA4108, Universit\'e de Rouen, 76821 Mont-Saint-Aignan Cedex, France \\ \email{\{Thierry.Lecroq,Arnaud.Lefebvre,Elise.Prieur\}@univ-rouen.fr} \and Dipartimento di Informatica, Universit\`a dell'Aquila, L'Aquila, Italy\\ \email{Filippo.Mignosi@di.univaq.it}}

\maketitle

\begin{abstract}
We investigate abelian repetitions in Sturmian words. We exploit a bijection between factors of Sturmian words and subintervals of the unitary segment that allows us to study the periods of abelian repetitions by using classical results of elementary Number Theory.  
If $k_{m}$ denotes the maximal exponent of an abelian repetition of period $m$, we prove that $\limsup k_{m}/m\ge \sqrt{5}$ for any Sturmian word, and the equality holds for the Fibonacci infinite word.
We further prove that 
the longest prefix of the Fibonacci infinite word that is an abelian repetition of period $F_j$, $j>1$, has length $F_j( F_{j+1}+F_{j-1} +1)-2$ if $j$ is even or $F_j( F_{j+1}+F_{j-1} )-2$ if $j$ is odd.
This allows us to give an exact formula for the smallest abelian periods of the Fibonacci finite words. More precisely, we prove that for $j\geq 3$, the Fibonacci word $f_j$  has abelian period equal to $F_n$, where  $n = \lfloor{j/2}\rfloor$ if $j = 0, 1, 2\mod{4}$, or $n = 1 + \lfloor{j/2}\rfloor$ if  $ j  = 3\mod{4}$.
\end{abstract}

\section{Introduction}

The study of repetitions in words is a classical subject in Theoretical Computer Science both from the combinatorial and the algorithmic point of view. Repetitions are strictly related to the notion of periodicity. Recall that a word $w$ of length $|w|$ has a \emph{period} $p>0$ if $w[i]=w[i+p]$ for any $1\leqslant i \leqslant |w|-p$, where $w[i]$ is the symbol in position $i$ of $w$. Every word $w$ has a minimal period $p\le |w|$. If $|w|/p\ge 1$, then $w$ is called a \emph{repetition} of period $p$ and \emph{exponent} $|w|/p$. When $|w|/p=k$ is an integer, the word $w$ is called an \emph{integer power}, since it can be written as $w=u^{k}$, i.e., $w$ is the concatenation of $k$ copies of a word $u$ of length $p$. If instead $|w|/p$ is not an integer, the word $w$ is called a \emph{fractional power}. So one can write $w=u^{k}v$, where $v$ is the prefix of $u$ such that $|w|/p=k+|v|/|u|$. For example, the word $w=aabaaba$ is a $7/3$-power since it has minimal period $3$ and length $7$. 
A classical reference on periodicity is \cite[Chap.~7]{lothaire-book:2002}.

Abelian properties concerning words have been studied since 
the very beginning of Formal Languages and Combinatorics on Words. The notion of Parikh vector  has become a standard and is often used without an explicit reference to the original 1966 Parikh's paper \cite{Parikh:1966:CL:321356.321364}. Abelian powers were first considered  in 1961 by Erd\"os \cite{Erdos1961221} as a natural generalization of usual powers.
Research concerning abelian 
properties of words and languages developed afterwards in 
different directions. 
In particular, there is a recent increasing of interest on 
abelian properties of words linked to periodicity (see, for example, 
\cite{AKP2012,CRSZ2010,DR2012,PZ13,Richomme201179,SS2011}), and on the algorithmic search of abelian periodicities in strings \cite{ChriCroIliop,Cro13,PSC11,PSC12,Stacs13}.
 
Recall that the Parikh vector $\PV_{w}$ of a finite word $w$ enumerates the cardinality of each letter of the alphabet in $w$. Therefore, two words have the same Parikh vector if one can be obtained from the other by permuting letters. 
We say that the word $w$ is an \emph{abelian repetition} of (abelian) period $m$ and exponent $|w|/m$ if $w$ can be written as $w=u_0u_1 \cdots u_{j-1}u_j$ for words $u_{i}$ and an integer $j>2$, where for $0<i<j$ all the $u_i$'s have the same Parikh vector $\PV$ whose sum of components is $m$ and the Parikh vectors of $u_0$ and $u_j$ are contained in $\PV$ (see \cite{CI2006}).  
When $u_{0}$ and $u_{j}$ are empty, $w$ is called an \emph{abelian power} or \emph{weak repetition}~\cite{Cummings_weakrepetitions}. For example, the word $w=abaab$ is an abelian repetition of period $2$, since one can set $u_{0}=a$, $u_{1}=ba$, $u_{2}=ab$ and $u_{3}=\varepsilon$, where $\varepsilon$ denotes the empty word.


It is well known that Sturmian words and Fibonacci words, in particular, are extremal cases for several problems related to repetitions (see for example \cite{CrIlRy09,IlMoSm97,Mignosi2012199}) and are worst-case examples for classical pattern matching algorithms, e.g. Knuth-Morris-Pratt \cite{aho:90,KoKu99}.
There exists a huge bibliography concerning Sturmian words (see for instance the survey papers 
\cite{BerstelRecent07,Berstel-Reutenauersurvey}, \cite[Chap. 2]{lothaire-book:2002}, 
\cite[Chap. 6]{Pitheasfogg}  and references therein). In particular, there is an analogous result to the one presented in this paper 
concerning classical repetitions in the
Fibonacci infinite word \cite{MignosiPirillo}.
In \cite{Mignosi_Infinite_words_with},
a bijection between factors of Sturmian 
words and subintervals of the unitary segment is described. 
We show in this paper that this bijection preserves abelian properties of
factors (see Proposition \ref{pro:main}). 
Therefore, we are able to apply techniques of Number Theory coupled with Combinatorics on Words to obtain our main results.
More precisely, if $k_{m}$ denotes the maximal exponent of an abelian repetition of period $m$, we prove that $\limsup k_{m}/m\ge \sqrt{5}$ for any Sturmian word, and the equality holds for the Fibonacci infinite word.

We further prove that for any Fibonacci number $F_j$, $j>1$, the longest prefix of the Fibonacci infinite word that is an abelian repetition of period $F_j$ has length $F_j( F_{j+1}+F_{j-1} +1)-2$ if $j$ is even or $F_j( F_{j+1}+F_{j-1} )-2$ if $j$ is odd (Theorem \ref{pro:longest_prefix}).
This allows us to give an exact formula for the smallest abelian periods of the Fibonacci finite words. More precisely, we prove, in Theorem \ref{The:8}, that for $j\geq 3$, the Fibonacci word $f_j$  has abelian period equal to $F_n$, where  $n = \lfloor{j/2}\rfloor$ if $j = 0, 1, 2\mod{4}$, or $n = 1 + \lfloor{j/2}\rfloor$ if  $ j  = 3\mod{4}$.

Due to space constraints the proofs are omitted, but they will be included in an upcoming full version of the paper.

\section{Preliminaries}

Let $\Sigma=\{a_{1},a_{2},\ldots ,a_{\sigma}\}$ be a finite ordered alphabet of cardinality $\sigma$ and $\Sigma^*$ the set of words over $\Sigma$. We denote by $|w|$ the length of the word $w$. We write $w[i]$ the $i$-th symbol of $w$ and $w[i\pp j]$ the factor of $w$ from the $i$-th symbol to the $j$-th symbol, with $1\leqslant i \leqslant j\leqslant |w|$.
We denote by $|w|_a$ the number of occurrences of the symbol $a\in\Sigma$ in the word $w$.

The \emph{Parikh vector} of a word $w$, denoted by $\PVW$, counts the occurrences of each letter of $\Sigma$ in $w$, i.e., $\PVW=(|w|_{a_{1}},\ldots,|w|_{a_{\sigma}})$. Given the Parikh vector $\PVW$ of a word $w$, we denote by $\PVW [i]$ its $i$-th component
and by $|\PVW|$ the sum of its components. Thus, for a word $w$ and $1\leqslant i\leqslant\sigma$, we have $\PVW [i]=|w|_{a_i}$ and $|\PVW|=\sum_{i=1}^{\sigma}\PVW[i]=|w|$. Finally, given two Parikh vectors $\PV,\QV$, we write $\PV\subset \QV$ if $\PV[i]\leqslant \QV[i]$ for every $1\leqslant i\leqslant \sigma$ and $|\PV|<|\QV|$.

Following \cite{CI2006}, we give the definition below.

\begin{definition}
\label{def-ap}
A word $w$ is an abelian repetition of period $m>0$ and exponent $|w|/m = k$ if one can write $w=u_0u_1 \cdots u_{j-1}u_j$ for some $j>2$ such that $\PV_{u_{0}}\subset \PV_{u_{1}}=\ldots =\PV_{u_{j-1}}\supset \PV_{u_{j}}$, and $|\PV_{u_{1}}|=\ldots =|\PV_{u_{j-1}}|=m$.

An abelian power is an abelian repetition in which $u_{0}=u_{j}=\varepsilon$.
\end{definition}
 
We call $u_0$ and $u_j$ the \emph{head} and the \emph{tail} of the abelian repetition, respectively. 
Notice that the length $t=|u_j|$ of the tail is uniquely determined by $h=|u_{0}|$, $m$ and $|w|$, namely $t=(|w|-h) \bmod m$.

\begin{example}
The word $w=abaababa$ is an abelian repetition of period $2$ and exponent $4$, since one can write $w=a\cdot ba\cdot ab \cdot ab \cdot a$. Notice that $w$ is also an abelian repetition of period $3$ and exponent $8/3$, since $w=\varepsilon \cdot aba\cdot aba \cdot ba$.
\end{example}

In the rest of the paper, when we refer to an abelian repetition of period $m$, we always suppose that $m$ is the minimal abelian period of $w$. 

\begin{remark}
We adopt the convention that an abelian repetition of exponent $k\geq 2$ has also exponent $k'$ for any real number $k'$ such that $2\leq k' \leq k$. This is a standard convention widely adopted in the classical case.
\end{remark}

\subsection{Sturmian words}

From now on, we fix the alphabet $\Sigma=\{\sa{a,b}\}$. We start by recalling a bijection between factors of Sturmian words and subintervals of the unitary segment introduced in 
\cite{Mignosi_Infinite_words_with}.

Let $\alpha$ and $\rho$ be two real numbers with $\alpha\in (0,1)$. Following the notations of 
\cite{Hardy_and_Wright}, the fractional part of a number $r$ is defined by $\{r\}=r-\lfloor r \rfloor$, where $\lfloor r \rfloor$ is the greatest integer smaller than or equal to $r$. Therefore, for $\alpha\in (0,1)$, one has that $\{-\alpha\}= 1-\alpha$.

\begin{figure}[!ht]
\centering
\fbox{
\includegraphics[scale=0.8]{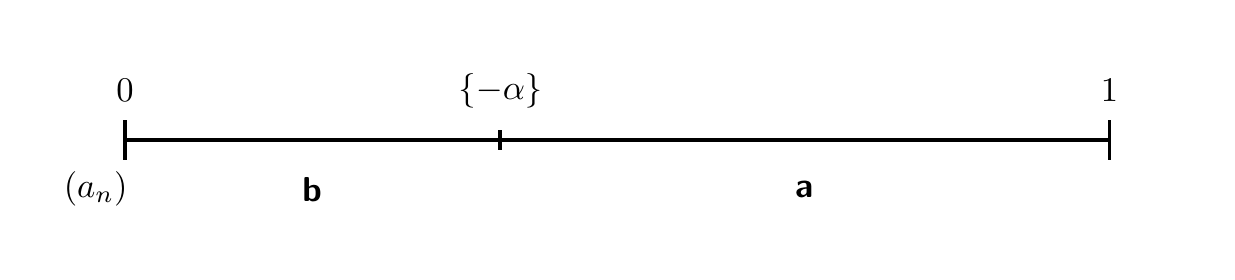} 
}\caption{An application of Proposition \ref{prima} when $\alpha=\phi-1\approx0.618$ (thus $\{-\alpha\}\approx0.382$) for $i=0$. If $\{n\alpha+\rho\} \in [\{-\alpha\},1)$, then $a_n=\sa{a}$; otherwise $a_n=\sa{b}$.}
\label{Fig:an}
\end{figure}

The sequence $\{ n\alpha +\rho \}, n > 0$, defines an infinite word $s_{\alpha,\rho} = a_1(\alpha,\rho)a_2(\alpha,\rho)\cdots$ by the rule
$$a_n(\alpha,\rho) =
\left\{
	\begin{array}{ll}
		\sa{b}  & \mbox{if } \{ n\alpha +\rho \}\in [0,\{-\alpha\}), \\
		\sa{a}  & \mbox{if } \{ n\alpha +\rho \}\in [\{-\alpha\}, 1).
	\end{array}
\right.$$ 
See \figurename~\ref{Fig:an} for a graphical illustration. 

We will write $a_n$ instead of $a_n(\alpha,\rho)$ whenever there is no possibility of mistake. If $\alpha$ is rational, i.e. $\alpha = n/m$, with $n$ and
$m$ coprime integers, then it is easy to prove that the word $s_{\alpha,\rho}$ is periodic and  $m$ is its minimal period. In this case, $s_{\alpha,\rho}$ is also periodic in the abelian sense, since it trivially has abelian period $m$.

If instead $\alpha$ is irrational, then $s_{\alpha,\rho}$ is not periodic and is called a \emph{Sturmian word}. Therefore, in the rest of the paper, we always suppose $\alpha$ irrational.

\begin{figure}[!ht]
\centering
\fbox{
\includegraphics[scale=0.8]{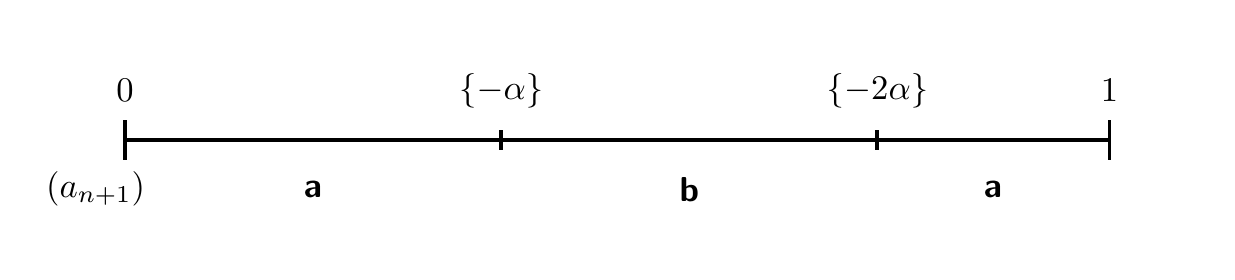} 
}\caption{An application of Proposition \ref{prima} when $\alpha=\phi-1\approx0.618$ (thus $\{-\alpha\}\approx0.382$) for $i=1$. If $\{n\alpha+\rho\} \in [0,\{-\alpha\})\cup [\{-2\alpha\},1)$, then $a_{n+1}=\sa{a}$; otherwise $a_{n+1}=\sa{b}$.}
\label{Fig:an+1senzaro}
\end{figure}

\begin{figure}[!ht]
\centering
\fbox{
\includegraphics[scale=0.8]{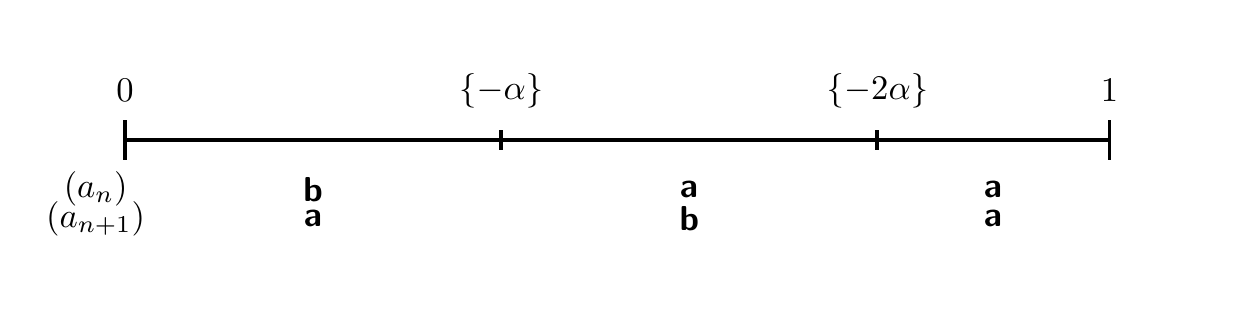}   
}\caption{A single graphic representation of the information given in \figurename~\ref{Fig:an} and ~\ref{Fig:an+1senzaro}. If $\{n\alpha +\rho\} \in [0,\{-\alpha\})=L_{0}(\alpha,2)$, then $a_{n}=\sa{b}$, $a_{n+1}=\sa{a}$. If $\{n\alpha +\rho\} \in [\{-\alpha\},\{-2\alpha\})=L_{1}(\alpha,2)$, then $a_{n}=\sa{a}$, $a_{n+1}=\sa{b}$. If $\{n\alpha +\rho\} \in [\{-2\alpha\},1)=L_{2}(\alpha,2)$, then $a_{n}=\sa{a}$, $a_{n+1}=\sa{a}$.}
\label{Fig:anan+1}
\end{figure}

\begin{example}
 For $\alpha= \phi-1$ and $\rho=0$, where $\phi=(1+\sqrt{5})/2$ is the golden ratio, one obtains the Fibonacci infinite word
$$f = \sa{abaababaabaababaababa} \cdots$$
\end{example}

\begin{remark}
 Since $\alpha \in (0,1)$, we have $\{-i\alpha\}\neq \{-(i+1)\alpha\}$ for any natural number $i$. We shall use this fact freely and with no explicit mention.
\end{remark}

It is possible to prove (see \cite[Corollary 2.3]{Mignosi_Infinite_words_with}) 
that the following result holds.

\begin{proposition}\label{prima}
Let $\alpha$ and $\rho$ be real numbers, with $\alpha\in (0,1)$ irrational. For any natural numbers $n$, $i$, with $n>0$, if $\{-(i+1)\alpha\}<\{-i\alpha\}$ then
$$ a_{n+i}=\sa{a} \iff \{ n\alpha +\rho \} \in [\{-(i+1)\alpha\},\{-i\alpha\}  ),$$ whereas if $\{-i\alpha\} < \{-(i+1)\alpha\})$ then
$$a_{n+i}=\sa{a} \iff \{ n\alpha +\rho \} \in [0,\{-i\alpha\}  ) \cup [\{-(i+1)\alpha\},1).$$
\end{proposition}

In \figurename~\ref{Fig:an} and \ref{Fig:an+1senzaro} we display a graphical representation of the formula given in  Proposition \ref{prima} for $\alpha=\phi-1$ when $i=0$ and $i=1$, respectively. In \figurename~\ref{Fig:anan+1} we present within a single graphic the situations illustrated in \figurename~\ref{Fig:an} and \ref{Fig:an+1senzaro}.

Let $m$ be a positive integer. Consider the $m+2$ points $0,1, \{-i\alpha\}$, for $1 \leq i\leq  m$. Rearranging these points in increasing order one has:
$$ 0=c_0(\alpha,m) <  c_1(\alpha,m) < \ldots < c_k(\alpha,m) < \ldots < c_m(\alpha,m) < c_{m+1}(\alpha,m)=1.$$
One can therefore define the $m+1$ non-empty subintervals
$$L_k(\alpha,m)=[c_k(\alpha,m), c_{k+1}(\alpha,m)),\ 0\leq k\leq m.$$
By using Proposition \ref{prima}, it is possible to associate with each interval $L_k(\alpha,m)$ a factor of length $m$ of the word $s_{\alpha,\rho}$, and this correspondence is bijective (see~\cite{Mignosi_On_the_number}). We call this correspondence the \emph{Sturmian bijection}.

\begin{proposition}\label{seconda}
Each factor of $s_{\alpha,\rho}$ of length $m$, $a_n a_{n+1} \cdots a_{n+m-1}$, depends only on the interval $L_k(\alpha,m)$  containing the point $\{n\alpha + \rho\}$; more precisely, it depends only on the set $I_k(\alpha,m)$ of integers $i\in \{0,1,\ldots , m-1\}$  such that either $\{-(i+1)\alpha\}<\{-i\alpha\}$ and $c_k(\alpha,m)\in [\{-(i+1)\alpha\},\{-i\alpha\})$ or $\{-(i+1)\alpha\}>\{-i\alpha\}$ and $c_k(\alpha,m) \notin [\{-i\alpha\},\{-(i+1)\alpha\})$. The set $I_k(\alpha,m)$ is the set of the integers $i$, with $0\leq i \leq m-1$, such that $a_{n+i}=\sa{a}$. 
\end{proposition}

\begin{corollary}\label{cor:rhozero}
Since the set of factors of $s_{\alpha,\rho}$ depends only on the 
sequence $\{-i \alpha\}$,  $i>0$, it does not depend on $\rho$. In particular, then, for any $\rho$ the word $s_{\alpha,\rho}$ has the same set of factors of the word $s_{\alpha,0}$.
\end{corollary}

\begin{example}
Let $\alpha=\phi-1$. In \figurename~\ref{Fig:anan+1} we show an example of the Sturmian bijection when $m=2$. The ordered sequence of points defining the subintervals $L_{k}(\alpha,2)$ is $$c_0(\alpha,2)=0, \ c_1(\alpha,2)=\{-\alpha\}\approx0.382, 
\ c_2(\alpha,2)=\{-2\alpha\}\approx0.764, \ c_3(\alpha,2)=1.$$

\begin{figure}[t]
\centering
\fbox{
\includegraphics[scale=0.9]{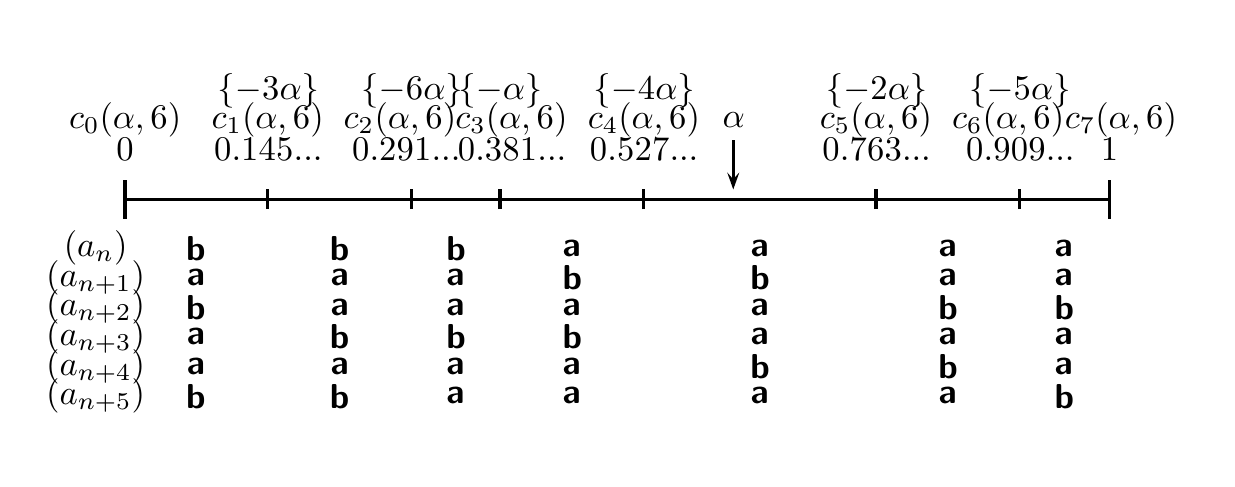}  
}\caption{The subintervals $L_k(\alpha,m)$ of the Sturmian bijection obtained for $\alpha=\phi-1$ 
and $m=6$. Below each interval there is
the factor of $s_\alpha$ of length $6$ associated with that interval. For $\rho=0$ and
$n=1$, the prefix of length $6$ of the Fibonacci word is associated 
with $L_4(\alpha,6)=[c_4(\alpha,6),c_5(\alpha,6))$, which is the interval 
 containing $\alpha$.}
\label{Fig:fattoridifibonacci6}
\end{figure}
\end{example}

In \figurename~\ref{Fig:fattoridifibonacci6} we show an example of the Sturmian bijection when $\alpha=\phi-1$ and $m=6$.  Below each interval there is the factor of $s_\alpha$ of length $m=6$ associated with that interval. The prefix of length $6$ of the Fibonacci word corresponds to the factor below the interval containing $\alpha$ (so, for $n=1$ and $\rho=0$). Notice that all the factors of length $6$ of the Fibonacci word appear, and moreover they are lexicographically ordered from right to left. This property concerning lexicographic order holds for any Sturmian word and any length $m$ of factors, and is stated in next proposition, which is of independent interest and is related to some recent research on Sturmian words and the lexicographic order (see \cite{Bucci201225,Glen200845,JeZa04,Perrin2012265}).

\begin{proposition}\label{Pro:otherbranch}
Let $m\geq 1$ and $k,k'$ such that $0\le k,k'\le m$. Then $k<k'$ if and only if the factor  $t_{\alpha,\rho,m}$
associated to  $L_k(\alpha,m)$ in the Sturmian bijection is lexicographically greater than the factor $t'_{\alpha,\rho,m}$
associated to $L_{k'}(\alpha,m)$.
\end{proposition}

In the next section we present a new property of the Sturmian bijection, that will allow us to use some standard Number Theory techniques to deal with abelian repetitions in Sturmian words and, in particular, in the Fibonacci infinite word.
Similar techniques are used in \cite{Richomme201179} to derive some other interesting results on abelian powers in Sturmian words.

\section{Sturmian bijection and Parikh vectors}

Let $s_{\alpha,\rho}$ be a Sturmian word. Since we are mainly interested in the set of factors of $s_{\alpha,\rho}$,  we do not lose generality, by Corollary \ref{cor:rhozero}, supposing $\rho=0$. 
The Sturmian words with $\rho=0$ are called \emph{characteristic}, and have been the object of deep studies within the field of Sturmian words. 
For simplicity of notation, we will write $s_\alpha$ instead of $s_{\alpha,0}$.

We now describe some properties of the Sturmian bijection between the factors of length $m$ of $s_{\alpha}$ and the subintervals $L_k(\alpha,m)$, that we will use to prove the main results of the paper.

\begin{proposition}\label{pro:main}
Under the Sturmian bijection, all the factors corresponding to an interval 
$c_k(\alpha,m)=[x,y)$ with $x\geq \{-m\alpha\}$ have the same Parikh 
vector $v_1(\alpha,m)$ and  all the factors 
corresponding to an interval $[x,y)$ with $y\leq \{-m\alpha\}$ 
have the same Parikh vector $v_2(\alpha,m)$. Moreover, one has $v_1(\alpha,m)[1] = v_2(\alpha,m)[1]+1$.
\end{proposition}

The reader can see in \figurename~\ref{Fig:fattoridifibonacci6} that the factors of length $6$ 
corresponding to an interval to the left of $\{-6(\phi-1)\}$ have Parikh vector $(3,3)$, while the other ones have Parikh vector $(4,2)$.

We now address the following  questions:

\begin{enumerate}
 \item Given $m$, how large can be the exponent of an abelian repetition of period $m$  in $s_\alpha$?
 \item What can we say in the particular case of the Fibonacci word, i.e., when $\alpha=\phi-1$?
\end{enumerate}

The next result follows straightforwardly from Proposition \ref{pro:main}.

\begin{corollary}\label{cor:2}
Let $w$ be an abelian power of period $m$ and exponent $k+1$  appearing in $s_\alpha$ in 
position $n$. Then all the points in the sequence 
$\{n\alpha\},\{(n+m)\alpha\},\{(n+2m)\alpha\},\ldots,\{(n+km)\alpha\}$ 
are in the same subinterval in which 
$[0,1)$ is subdivided by the point $ \{-m\alpha\}$, 
i.e., either $[0,\{-m\alpha\})$ or $[\{-m\alpha\},1)$.
\end{corollary}

The next proposition is a technical step to prove the following theorem.

\begin{proposition}
If $k\geq 1$, the $k+1$ points of Corollary \ref{cor:2} are naturally ordered. That is to say, if $\{m\alpha\} < 0.5$, then they are all in the subinterval 
$[0, \{-m\alpha\})$ and one has $\{n\alpha\}<\{(n+m)\alpha\}< \ldots < \{(n+km)\alpha\}$; if instead $\{m\alpha\}>0.5$ then they are all in the interval $[\{-m\alpha\},1)$ 
and one has $\{(n+km)\alpha\} < \{(n+(k-1)m)\alpha\} < \ldots < \{n\alpha\} $.
\end{proposition}

\begin{theorem}\label{the:main} 
Let $m$ be a positive integer such that $\{m\alpha\}< 0.5$ (resp. $\{m\alpha\}> 0.5)$. Then:
\begin{enumerate}
 \item In $s_\alpha$ there is an abelian power of period $m$ and exponent $k\geq 2$ if and only if $\{m\alpha\}< \frac{1}{k}$ (resp.~$\{-m\alpha\}< \frac{1}{k}$).
 
 \item If in $s_\alpha$ there is an abelian power of period $m$ and exponent $k\geq 2$ starting in position $i$ with $\{i\alpha\}\geq\{m\alpha\}$ (resp. $\{i\alpha\}\leq\{m\alpha\}$),
then $\{m\alpha\}< \frac{1}{k+1}$ (resp. $\{-m\alpha\}< \frac{1}{k+1}$ ).
Conversely, if  $\{m\alpha\}< \frac{1}{k+1}$  (resp. $\{-m\alpha\}< \frac{1}{k+1}$), then there is an abelian power of period $m$ and exponent $k\geq 2$ starting in position $m$.
\end{enumerate}
\end{theorem}

The previous theorem allows us to deal with abelian repetitions in a Sturmian word $s_{\alpha}$ by using classical results on the approximation of the irrational $\alpha$ by rationals. This is a classical topic in Number Theory. Since the number $\phi -1$ has special properties within this topic, we have in turn specific results for the Fibonacci infinite word.

\section{Approximating irrationals by rationals and abelian repetitions}

We recall some classical results of Number Theory. For any notation not explicitly defined in this section we refer to \cite[Chap.~X, XI]{Hardy_and_Wright}.

The sequence $F_0=1, F_1=1, F_{j+1}=F_j+F_{j-1}$ for $j\geq 1$  is the well known sequence of Fibonacci numbers. The sequence of fractions $\frac{F_{j+1}}{F_j}$ converges to $\phi=\frac{\sqrt{5}+1}{2}$, while the sequence $\frac{F_{j}}{F_{j+1}}$ converges to $\phi-1=\frac{\sqrt{5}-1}{2}$. Moreover, the sequences $\frac{F_{j+1}}{F_j}$ and $0=\frac{0}{1}, $
$\frac{F_{j}}{F_{j+1}}$, $j= 0, 1, \ldots$, are the sequences of convergents, in the development in continued fractions, of $\phi$ and $\phi-1$ respectively.
 
Concerning the approximation given by the above convergents, the following result holds (see \cite[Chap.~X, Theorem 171]{Hardy_and_Wright} and \cite[Chap.~XI, Section 11.8]{Hardy_and_Wright}).
 
\begin{theorem}\label{the:jmeno1}
For any $j>0$, $$\phi-\frac{F_{j+1}}{F_j}=(\phi-1)-\frac{F_{j-1}}{F_j}=\frac{(-1)^{j}}{F_j(\phi F_j+F_{j-1})}.$$
\end{theorem}
 
We also report the following theorems (see \cite[Chap.~XI, Theorem 193 and the proof of Theorem 194]{Hardy_and_Wright}). 

\begin{theorem}\label{theor:appr}
Any irrational $\alpha$ has an infinity of approximations which satisfy
$$\left|\frac{n}{m}-\alpha\right| < \frac{1}{\sqrt{5}m^2}.$$
\end{theorem}

\begin{theorem}\label{the:fibo1}
Let $\alpha=\phi-1$. If $A>\sqrt{5}$, then the inequality 
$$\left|\frac{n}{m}-\alpha\right|< \frac{1}{Am^2}$$
has only a finite number of solutions.
\end{theorem}

The last two theorems, coupled with the first part of Theorem \ref{the:main}, allow us to derive the next result. 

\begin{theorem}\label{theor:sqrt5}
Let $s_{\alpha}$ be a Sturmian word. For any integer $m>1$, let $k_{m}$ be the maximal exponent of an abelian repetition of period $m$ in $s_{\alpha}$. Then $$\limsup_{m \to \infty} \frac{k_{m}}{m}\geq \sqrt{5},$$ and the equality holds if $\alpha=\phi-1$.
\end{theorem}

\section{Prefixes of the Fibonacci infinite word}

We now study the abelian repetitions that are prefixes of the Fibonacci infinite word. For this, we will make use of the second part of 
Theorem \ref{the:main}.
Notice that an abelian repetition of period $m$ appearing as a prefix of the Fibonacci word can have a head of length equal to $m-1$ at most. Therefore, we have to check all the abelian powers that start in position $i$ for every $i= 1,\ldots, m$. In order to do this, we report here another result (see \cite[Chap.~X, Theorem 182]{Hardy_and_Wright}).

\begin{theorem}\label{the:closest}
Let $n_i/m_i$ be the $i$-th convergent to $\alpha$. If $i>1$, $0< m\leq m_i$ and 
$n/m \neq n_i/m_i$, then $ |n_i-m_i\alpha|<|n-m\alpha|$.
\end{theorem}

The previous theorem implies the following result.
 
\begin{corollary}\label{cor:cor}
Suppose that  $m>1$ is the denominator of a convergent to $\alpha$ and that 
$\{m\alpha\}< 0.5$ (resp. $\{m\alpha\}> 0.5)$. Then for any $i$ such that $1\leq i <m$, one has 
$\{i\alpha\}\geq\{m\alpha\}$ (resp.~$\{i\alpha\}\leq\{m\alpha\}$).
\end{corollary}

From the previous corollary, we have that if $m>1$ is a Fibonacci number and $\alpha =\phi-1$, then the hypotheses of the second part of Theorem \ref{the:main} are satisfied. The next proposition is a direct consequence of Corollary \ref{cor:cor}, Theorem \ref{the:main}
and Theorem \ref{the:jmeno1}.

\begin{proposition}\label{pro:maximal_exponent}
Let $j>1$. In the Fibonacci infinite word, the longest abelian power having period $F_j$ and starting in a position $i\leq F_j$ has an occurrence starting in position $F_j$, and has exponent equal to
$$\lfloor \phi F_j+F_{j-1} \rfloor -1=
\begin{cases}
 F_{j+1}+F_{j-1} -1 &  \mbox{ if $j$ is even;}\\
 F_{j+1}+F_{j-1} -2 &  \mbox{ if $j$ is odd.}
\end{cases}
 $$
\end{proposition}

The following theorem provides a formula for computing the length of the longest abelian repetition occurring as a prefix in the Fibonacci infinite word.

\begin{theorem}\label{pro:longest_prefix}
Let $j>1$. The longest prefix of the Fibonacci infinite word that is an abelian repetition of period $F_j$ has length $F_j( F_{j+1}+F_{j-1} +1)-2$ if $j$ is even or $F_j( F_{j+1}+F_{j-1} )-2$ if $j$ is odd.
\end{theorem}

\begin{corollary}\label{corr:last}
Let $j>1$ and $k_{j}$ be the maximal exponent of a prefix of the Fibonacci word that is an abelian repetition of period $F_{j}$. Then $$\lim_{j\to \infty}\frac{k_{j}}{F_{j}}=\sqrt{5}.$$
\end{corollary}

\begin{figure}[t!]
\centering
\fbox{
\includegraphics[scale=0.8]{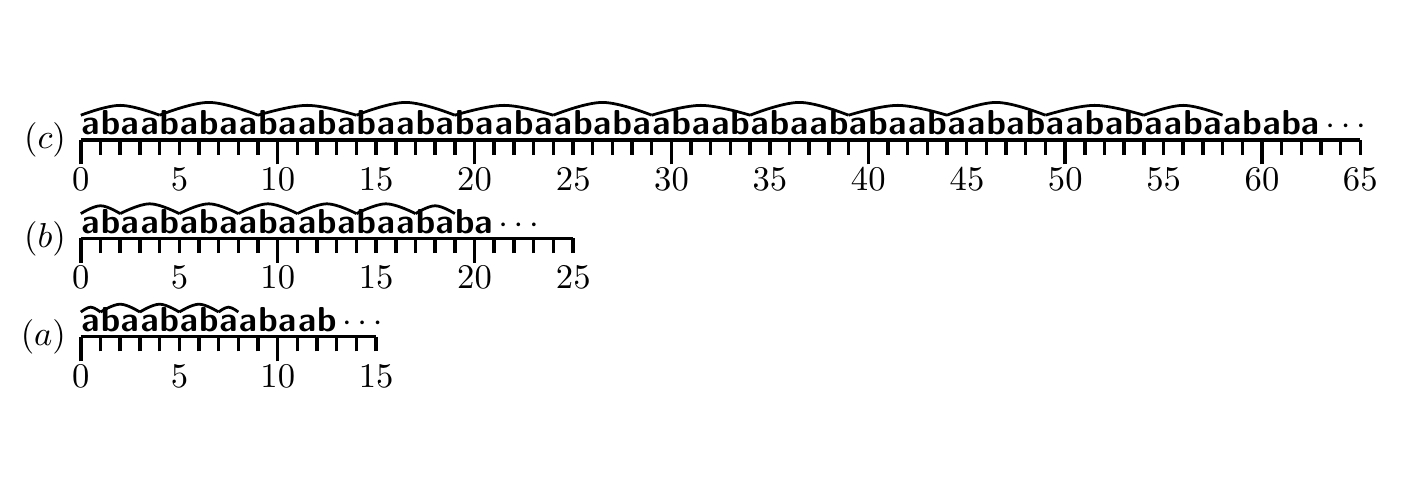} 
}\caption{Longest abelian repetition of period $m$ that is a  prefix of the  Fibonacci word for $m=2,3,5$. 
$(a)$ For $m=2$, the longest abelian repetition has length $8=1+3p+1$. 
$(b)$ For $m=3$, the longest abelian repetition has length $19=2+5p+2$.
$(c)$ For $m=5$, the longest abelian repetition has length $58=4+10p+4$.}
\label{Fig:prefixlenghts}
\end{figure}

In \figurename~\ref{Fig:prefixlenghts} we give a graphical representation of the longest prefix of the Fibonacci infinite word that is an abelian repetition of period $m$ for $m=2,3$ and $5$. In Table \ref{tab:fici} we give the length $\lp(F_{j})$ of the longest prefix of the Fibonacci infinite word that is an abelian repetition of period $F_{j}$, for $j=2,\ldots, 11$, computed using the formula of Theorem \ref{pro:longest_prefix}. We also show the values of the distance between $\sqrt{5}$ and the ratio between the maximal exponent $k_{j}=\lp(F_{j})/F_{j}$ of a prefix of the Fibonacci infinite word having abelian period $F_{j}$ and $F_{j}$.

\begin{table}[bt]
\centering  
\begin{scriptsize}
\begin{raggedright}
\begin{tabular}{c *{30}{@{\hspace{3.1mm}}l}}
$j$\hspace{2mm} & 2& 3& 4& 5& 6& 7& 8& 9& 10& 11 
\\[1mm]
\hline \\
$F_{j}$\hspace{2mm} & 2& 3& 5& 8& 13& 21& 34& 55& 89& 144 
\\[1mm]
\hline \\
$\lp(F_{j})$ \hspace{2mm}   & 8 & 19  & 58  & 142  & 388  & 985  & 2616\hspace{1ex} & 6763  & 17798 & 46366 
\\[1mm]
 \hline\\
$|\sqrt{5}-k_{j}/F_{j}|\times 10^2$ \hspace{2mm}   & $23.6 $ & $12.5 $ & $8.393 $ & $1.732 $ & $5.98 $ & $0.25 $ & $2.69 $ & $0.037 $ & $1.087$ & $0.005 $ 
\\[1mm]
\hline \rule[0pt]{0pt}{12pt}
\end{tabular}
\end{raggedright}\caption{\label{tab:fici}The length of the longest prefix ($\lp(F_{j})$) of the Fibonacci word having abelian period $F_{j}$ for $j=2,\ldots, 11$.  The table also reports rounded distances (multiplied by $10^2$) between $\sqrt{5}$ and the ratio between the exponent $k_{j}=\lp(F_{j})/F_{j}$ of the longest prefix of the Fibonacci word having abelian period $F_{j}$ and $F_{j}$ (see Corollary~\ref{corr:last}).}
\end{scriptsize}
\end{table}

Recall that the Fibonacci (finite) words are defined by $f_{0}=\sa{b}$, $f_{1}=\sa{a}$, and for every $j>1$, $f_{j+1}=f_{j}f_{j-1}$. So, for every $j$, one has $|f_{j}|=F_{j}$.
As a consequence of the formula given in Theorem \ref{pro:longest_prefix}, we have the following result on the smallest abelian periods of the Fibonacci words.

\begin{theorem}\label{The:8} 
For $j \geq 3$, the (smallest) abelian period of the word $f_j$ is the $n$-th Fibonacci number $F_n$, where  $n = \lfloor{j/2}\rfloor$ if $j = 0, 1, 2\mod{4}$, or $n = 1 + \lfloor{j/2}\rfloor$ if  $ j  = 3\mod{4}$. 
\end{theorem}

For example, the abelian period of the word $f_{4}=\sa{abaab}$ is $2=F_{2}=\lfloor{4/2}\rfloor$, since one can write $f_{4}=\sa{a}\cdot \sa{ba} \cdot \sa{ab}$; the abelian period of $f_{5}=\sa{abaababa}$ is $2=F_{2}$; the abelian period of $f_{6}=\sa{abaababaabaab}$ is $3=F_{3}$; the abelian period of $f_{7}=\sa{abaababaabaababaababa}$ is $5=F_{4}$.
In Table~\ref{tab:val} we report the abelian periods of the first Fibonacci words. 

\begin{table}[htb]
\centering  
\begin{small}
\begin{raggedright}
\begin{tabular}{c *{30}{@{\hspace{2.1mm}}l}}
 $j$\hspace{2mm}  & 3\hspace{1ex} & 4\hspace{1ex} & 5\hspace{1ex} & 6\hspace{1ex} & 7\hspace{1ex} & 8\hspace{1ex} & 9\hspace{1ex} & 10 & 11 & 12 & 13 & 14 & 15 & 16 \\
 \hline \rule[-6pt]{0pt}{22pt}
a.~p.~of $f_{j}$\hspace{2mm}  & $F_{2}$ & $F_{2}$ & $F_{2}$ & $F_{3}$ & $F_{4}$ & $F_{4}$ & $F_{4}$ & $F_{5}$ & $F_{6}$ & $F_{6}$ & $F_{6}$ & $F_{7}$ & $F_{8}$ & $F_{8}$\\
\hline \rule[-6pt]{0pt}{22pt}
\end{tabular}
\end{raggedright}\caption{\label{tab:val}The (smallest) abelian periods of the Fibonacci words $f_{j}$ for $j=3,\ldots, 16$.}
\end{small}
\end{table}

We conclude the paper with the following open problems:

\begin{enumerate}
\item 
Is it possible to find the exact value of $\limsup \frac{k_m}{m}$ for other Sturmian words $s_\alpha$ with slope $\alpha$ different from $\phi -1$? 
\item 
Is it possible to give the exact value of this superior limit when
$\alpha$ is an algebraic number of degree $2$?  
\end{enumerate}

\bibliographystyle{splncs}
\bibliography{ref}

\begin{thebibliography}{10}

\bibitem{lothaire-book:2002}
Lothaire, M.:
\newblock Algebraic Combinatorics on {W}ords.
\newblock Cambridge University Press, Cambridge, U.K. (2002)

\bibitem{Parikh:1966:CL:321356.321364}
Parikh, R.J.:
\newblock On context-free languages.
\newblock J. Assoc. Comput. Mach. \textbf{13} (1966)  570--581

\bibitem{Erdos1961221}
Erd\"os, P.:
\newblock Some unsolved problems.
\newblock Magyar Tud. Akad. Mat. Kutato. Int. Kozl. \textbf{6} (1961)  221--254

\bibitem{AKP2012}
Avgustinovich, S., Karhum{\"a}ki, J., Puzynina, S.:
\newblock On abelian versions of {C}ritical {F}actorization {T}heorem.
\newblock RAIRO Theor. Inform. Appl. \textbf{46} (2012)  3--15

\bibitem{CRSZ2010}
Cassaigne, J., Richomme, G., Saari, K., Zamboni, L.:
\newblock Avoiding {A}belian powers in binary words with bounded {A}belian
  complexity.
\newblock Int. J. Found. Comput. Sci. \textbf{22} (2011)  905--920

\bibitem{DR2012}
Domaratzki, M., Rampersad, N.:
\newblock Abelian primitive words.
\newblock Int. J. Found. Comput. Sci. \textbf{23} (2012)  1021--1034

\bibitem{PZ13}
Puzynina, S., Zamboni, L.Q.:
\newblock Abelian returns in {S}turmian words.
\newblock J. Comb. Theory, Ser. A \textbf{120} (2013)  390--408

\bibitem{Richomme201179}
Richomme, G., Saari, K., Zamboni, L.:
\newblock Abelian complexity of minimal subshifts.
\newblock Journal of the London Mathematical Society \textbf{83} (2011)  79--95

\bibitem{SS2011}
Samsonov, A., Shur, A.:
\newblock On {A}belian repetition threshold.
\newblock RAIRO Theor. Inform. Appl. \textbf{46} (2012)  147--163

\bibitem{ChriCroIliop}
Christou, M., Crochemore, M., Iliopoulos, C.S.:
\newblock Identifying all abelian periods of a string in quadratic time and
  relevant problems.
\newblock Int. J. Found. Comput. Sci. \textbf{23} (2012)  1371--1384

\bibitem{Cro13}
Crochemore, M., Iliopoulos, C.S., Kociumaka, T., Kubica, M., Pachocki, J.,
  Radoszewski, J., Rytter, W., Tyczynski, W., Walen, T.:
\newblock A note on efficient computation of all abelian periods in a string.
\newblock Inf. Process. Lett. \textbf{113} (2013)  74--77

\bibitem{PSC11}
Fici, G., Lecroq, T., Lefebvre, A., Prieur-Gaston, E.:
\newblock Computing {A}belian {P}eriods in {W}ords.
\newblock In: Proceedings of the Prague Stringology Conference, {PSC} 2011,
  Czech Technical University in Prague (2011)  184--196

\bibitem{PSC12}
Fici, G., Lecroq, T., Lefebvre, A., Prieur-Gaston, E., Smyth, W.F.:
\newblock {Q}uasi-{L}inear {T}ime {C}omputation of the {A}belian {P}eriods of a
  {W}ord.
\newblock In: Proceedings of the Prague Stringology Conference, {PSC} 2012,
  Czech Technical University in Prague (2012)  103--110

\bibitem{Stacs13}
Kociumaka, T., Radoszewski, J., Rytter, W.:
\newblock Fast algorithms for abelian periods in words and greatest common
  divisor queries.
\newblock In: STACS 2013. Volume~20 of LIPIcs., Schloss Dagstuhl -
  Leibniz-Zentrum fuer Informatik (2013)  245--256

\bibitem{CI2006}
Constantinescu, S., Ilie, L.:
\newblock {F}ine and {W}ilf's theorem for abelian periods.
\newblock Bull. Eur. Assoc. Theoret. Comput. Sci. EATCS \textbf{89} (2006)
  167--170

\bibitem{Cummings_weakrepetitions}
Cummings, L.J., Smyth, W.F.:
\newblock Weak repetitions in strings.
\newblock {J. Combin. Math. Combin. Comput.} \textbf{24} (1997)  33--48

\bibitem{CrIlRy09}
Crochemore, M., Ilie, L., Rytter, W.:
\newblock Repetitions in strings: {A}lgorithms and combinatorics.
\newblock Theoret. Comput. Sci. \textbf{410} (2009)  5227--5235

\bibitem{IlMoSm97}
Iliopoulos, C.S., Moore, D., Smyth, W.F.:
\newblock A {C}haracterization of the {S}quares in a {F}ibonacci {S}tring.
\newblock Theoret. Comput. Sci. \textbf{172} (1997)  281--291

\bibitem{Mignosi2012199}
Mignosi, F., Restivo, A.:
\newblock Characteristic {S}turmian words are extremal for the critical
  factorization theorem.
\newblock Theoret. Comput. Sci. \textbf{454} (2012)  199 -- 205

\bibitem{aho:90}
Aho, A.:
\newblock {Algorithms for Finding Patterns in Strings}.
\newblock In van Leeuwen, J., ed.: Handbook of Theoret. Comput. Sci.
\newblock Elsevier Science Publishers B. V., Amsterdam, the Netherlands (1990)
  257--300

\bibitem{KoKu99}
Kolpakov, R., Kucherov, G.:
\newblock Finding {M}aximal {R}epetitions in a {W}ord in {L}inear {T}ime.
\newblock In: Proceedings of the 40th Annual Symposium on Foundations of
  Computer Science. FOCS '99, IEEE Computer Society (1999)  596--604

\bibitem{BerstelRecent07}
Berstel, J.:
\newblock {S}turmian and episturmian words (a survey of some recent results).
\newblock In Bozapalidis, S., Rohonis, G., eds.: CAI 2007. Volume 4728 of
  Lecture Notes in Comput. Sci., Springer (2007)  23--47

\bibitem{Berstel-Reutenauersurvey}
Berstel, J., Lauve, A., Reutenauer, C., Saliola, F.:
\newblock Combinatorics on Words: Christoffel Words and Repetition in Words.
  Volume~27 of CRM monograph series.
\newblock American Mathematical Society (2008)

\bibitem{Pitheasfogg}
Pytheas~Fogg, N.:
\newblock Substitutions in Dynamics, Arithmetics and Combinatorics. Volume 1794
  of Lecture Notes in Math.
\newblock Springer (2002)

\bibitem{MignosiPirillo}
Mignosi, F., Pirillo, G.:
\newblock Repetitions in the {F}ibonacci infinite word.
\newblock RAIRO Theor. Inform. Appl. \textbf{26} (1992)  199--204

\bibitem{Mignosi_Infinite_words_with}
Mignosi, F.:
\newblock Infinite {W}ords with {L}inear {S}ubword {C}omplexity.
\newblock Theoret. Comput. Sci. \textbf{65} (1989)  221--242

\bibitem{Hardy_and_Wright}
Hardy, G.H., Wright, E.M.:
\newblock An Introduction to the Theory of Numbers.
\newblock Clarendon Press, Oxford (1979) 5th edition.

\bibitem{Mignosi_On_the_number}
Mignosi, F.:
\newblock On the number of factors of {S}turmian words.
\newblock Theoret. Comput. Sci. \textbf{82} (1991)  71--84

\bibitem{Bucci201225}
Bucci, M., De~Luca, A., Zamboni, L.:
\newblock Some characterizations of {S}turmian words in terms of the
  lexicographic order.
\newblock Fundamenta Informaticae \textbf{116} (2012)  25--33

\bibitem{Glen200845}
Glen, A., Justin, J., Pirillo, G.:
\newblock Characterizations of finite and infinite episturmian words via
  lexicographic orderings.
\newblock European Journal of Combinatorics \textbf{29} (2008)  45--58

\bibitem{JeZa04}
Jenkinson, O., Zamboni, L.Q.:
\newblock Characterisations of balanced words via orderings.
\newblock Theoret. Comput. Sci. \textbf{310} (2004)  247--271

\bibitem{Perrin2012265}
Perrin, D., Restivo, A.:
\newblock A note on {S}turmian words.
\newblock Theoret. Comput. Sci. \textbf{429} (2012)  265--272

\end{thebibliography}

\end{document}